\documentclass[aoas,nameyear,seceqn,dvips]{arximspdf}
\usepackage{graphics}
\doi{10.1214/09-AOAS316}
\volume{4}
\issue{2}
\pubyear{2010}
\firstpage{962}
\lastpage{987}

\begin{document}
\begin{frontmatter}

\title{A nested mixture model for protein identification using mass
spectrometry}
\runtitle{A nested mixture model for protein identification}

\begin{aug}
\author[a]{\fnms{Qunhua} \snm{Li}\ead[label=e1]{qli@stat.washington.edu}\corref{}},
\author[b]{\fnms{Michael J.} \snm{MacCoss}\ead[label=e2]{maccoss@u.washington.edu}}
\and
\author[c]{\fnms{Matthew} \snm{Stephens}\ead[label=e3]{mstephens@uchicago.edu}}
\runauthor{Q. Li, M. J. M\textsc{ac}coss and M. Stephens}
\pdfauthor{Q. Li, M. Maccoss and M. Stephens}

\affiliation{University of Washington, University of Washington and
University of Chicago}

\address[a]{Q. Li\\ Department of Statistics\\ University of Washington\\
Box 354322 \\
Seattle, Washington 98195-4322\\ USA\\
\printead{e1}}

\address[b]{M. J. MacCoss\\ Department of Genome Sciences\\ University of
Washington\\Box 355065\\
Seattle, Washington 98195-5065\\ USA\\
\printead{e2}}

\address[c]{M. Stephens\\ Department of Statistics and Human Genetics\\
University of Chicago\\
Eckhart Hall Room 126\\
5734 S. University Avenue\\
Chicago, Illinois 60637\\ USA\\
\printead{e3}}
\end{aug}

% HISTORY:
\received{\smonth{10} \syear{2008}}
\revised{\smonth{9} \syear{2009}}

% ABSTRACT
%
\begin{abstract}
Mass spectrometry provides a high-throughput
way to identify proteins in biological samples. In a typical experiment,
proteins in a sample are first broken into their constituent peptides.
The resulting mixture
of peptides is then subjected to mass spectrometry, which generates
thousands of spectra, each
characteristic of its generating peptide. Here we consider the problem
of inferring, from these spectra, which proteins and peptides
are present in the sample.
We develop a statistical
approach to the problem, based on a nested mixture model. In contrast
to commonly used
two-stage approaches, this model provides a one-stage solution that
simultaneously identifies
which proteins are present, and which peptides are correctly
identified. In this way our model incorporates the
evidence feedback between proteins and their constituent peptides.
Using simulated data and a yeast data set, we compare and contrast our
method with
existing widely used approaches (PeptideProphet/ProteinProphet) and
with a recently published new approach,
HSM. For peptide identification, our single-stage approach yields consistently
more accurate results. For protein identification the methods have similar
accuracy in most settings, although we exhibit some scenarios in which
the existing methods
perform poorly.
\end{abstract}

\begin{keyword}
\kwd{Mixture model}
\kwd{nested structure}
\kwd{EM algorithm}
\kwd{protein identification}
\kwd{peptide identification}
\kwd{mass spectrometry}
\kwd{proteomics}.
\end{keyword}

\end{frontmatter}

%s1 ###
\section{Introduction}

\begin{figure}

\includegraphics{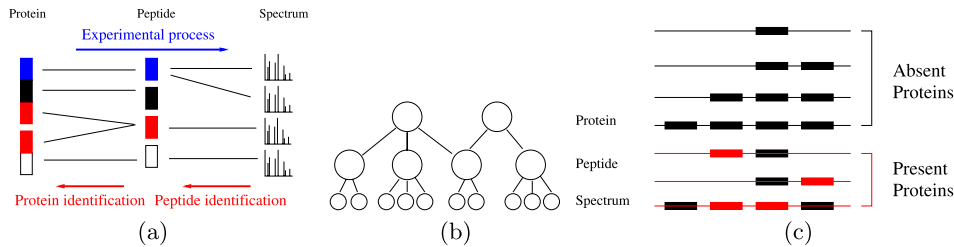}

\caption{\textup{(a)} Protein identification using mass spectrometry. Proteins
(left) are broken into constituent peptides (center), which are then
subjected to mass spectrometry to produce spectra (right). The
inference problem considered here is to infer which peptides, belonging
to which proteins, generated the observed spectra. \textup{(b)} Graphical
representation of the nested relationship between spectra, peptides and
proteins. \textup{(c)} Examples of putative protein identifications
reconstructed from putative peptide identifications. Proteins that are
truly absent from the sample will contain all incorrectly identified
peptides (black). Proteins that are present in the sample will
typically contain a mixture of correctly (red) and incorrectly (black)
identified peptides.}\label{F:flowchart}
\end{figure}

Protein identification using tandem mass spectrometry (MS/MS) is the
most widely used tool for identifying proteins in complex biological
samples  [\citet{steen04}]. In a typical MS/MS experiment [Figure~\ref
{F:flowchart}(a)], proteins in a sample are first broken into short
sequences, called peptides, and the resulting mixture of peptides is
subjected to mass spectrometry to generate tandem mass spectra, which
contains sequence information that is characteristic of its generating
peptide [\citet{coon05}; \citet{kinter00}]. The peptide that is most likely to
generate each spectrum then is identified using some computational
methods, for example, by matching to a list of theoretical spectra of
peptide candidates.
From these putative peptide identifications, the proteins that are
present in the mixture are then identified.
The protein identification problem is challenging, primarily because
the matching of spectra to peptides is highly error-prone:
80--90\% of identified peptides may be incorrect identifications if no
filtering is applied  [\citet{keller02b, nesvizhskii04}]. In
particular, to minimize errors in protein identifications, it is
critical to assess, and take proper account of, the strength of the
evidence for each putative peptide identification.

%f1 ###

Here we develop a statistical
approach to this problem, based on a nested mixture model.
Our method differs from most
previous approaches to the problem in that it is based on a {\it
single} statistical model
that incorporates latent variables indicating which proteins are present,
and which peptides are correctly identified.
Thus, instead of taking the more common sequential
approach to the problem (spectra $\rightarrow$ peptides $\rightarrow$
proteins), our model \textit{simultaneously}
estimates which proteins are present,
and which peptides are correctly identified, allowing for appropriate
evidence feedback between proteins and their constituent peptides.
This not only provides the potential for more accurate identifications
(particularly
at the peptide level), but, as we illustrate here, it also allows for
better calibrated estimates of uncertainty
in which identifications are correct. As far as we are aware, the only
other published method that takes a single-stage approach to the
problem is that of \citet{shen08}. Although Shen et al.'s
model shares the goal of our approach of allowing evidence feedback
from proteins to peptides, the structure of their model is quite
different from ours (see \hyperref[S:discussion]{Discussion} for more details), and, as we see
in our comparisons, the empirical performance of the methods can also
differ substantially.

In general statistical terms this problem involves a nested structure
of a form
that is encountered in other statistical inference problems (e.g.,
multilevel latent class models  [\citet{vermunt03}] and hierarchical
topic models  [\citet{blei04}]). These problems usually share two
common features: (1) there exists a physical or latent hierarchical
relationship between lower-level and upper-level elements; and
(2) only the lowest-level elements in the hierarchy are typically
observed. Here the nested structure is due to the subsequent
relationship between lower-level elements (peptides) and upper-level
elements (proteins) [Figure~\ref{F:flowchart}(b)]. The goals of inference will, of
course, vary depending on the application.
In this case the primary goal is to infer the states (i.e., presence or
absence in the mixture) of the upper-level elements, though the states
of the lower-level elements are also of interest.

The structure of the paper is as follows. Section \ref{S:methods}
describes the problem in more detail, reviews existing approaches, and
describes our modeling approach. Section \ref{S:Results} shows
empirical comparisons of our method with different approaches on both
real and simulated data. In Section \ref{S:discussion} we conclude and discuss
potential future enhancements.

%s2 ###
\section{Methods and models}\label{S:methods}

The first step in analysis of MS/MS data is typically to identify, for
each spectrum produced,
the peptide that is most likely to have generated the observed spectrum,
and to assign each such identification a score that reflects the
strength of the
evidence for the identification being correct. Often this process is
performed by searching
a database of potential peptides, and computing some measure of the
similarity between the observed spectrum and a
theoretical ``expected'' spectrum for each peptide in the database [\citet{sadygov04a}]. For each spectrum the
highest-scoring peptide is then reported, together with its score. Here
we assume
that this process has already been performed, and tackle the protein
identification problem:
using the list of putative peptides, and scores, to infer a list of
proteins that are likely to be
present in the mixture. Other important goals include accurately
assessing confidence for each protein identification, and inferring
which of the initial putative peptide identifications are actually correct.

%s2.1 ###
\subsection{Existing approaches}

Almost all current approaches to protein identification follow a
two-stage strategy:
\begin{enumerate}[1.]
\item The peptide identification scores are processed, together with
other relevant information (e.g., sequence
characteristics) on the identified peptide, to compute a statistical
measure of the strength of evidence for each peptide identification.
Although several methods exist [e.g., \citet{sadygov03}; \citet{kall07}], by far
the most widely used approach appears to be PeptideProphet [\citet{keller02}], which uses a mixture model to cluster the identified
peptides into correct and incorrect identifications, and to assign a
probability to each peptide
identification being correct.
\item The statistical measures of support for each peptide
identification are taken as input to a protein inference procedure.
These procedures infer the presence or absence of each protein, either
by simple ad hoc thresholding rules, for example, identifying proteins
as present if they contain two or more peptides with strong support, or
by more sophisticated means (ProteinProphet [\citet{nesvizhskii03}],
Prot\_Probe [\citet{sadygov04}] and EBP [\citet{price07}]). The basic
idea of ProteinProphet [\citet{nesvizhskii03}],
which is the most widely used of these methods, will be described below.
\end{enumerate}

This two-stage approach, although widely used, is sub-optimal. In
particular, it does
not allow for evidence to feed back, from the presence/absence status
of a protein to the status of its constituent peptides, as it should
due to the nested relationship between a protein and its peptides.
\citet{shen08} also note this problem with the two-stage
approach, and propose an alternative one-stage approach using a
latent-variable-based model.
Their model differs from ours in several aspects (see discussion), and
performs less well than our approach in the limited empirical
comparisons we consider here (see results).

%s2.2 ###
\subsection{A nested mixture model}\label{SS:nested}

The data consist of a large number of putative peptide identifications,
each corresponding to a single MS/MS spectrum, and each having a score
that relates to the strength of the evidence for the identification
being correct (higher scores corresponding to stronger evidence).
From this list of putative peptides, it is straightforward to
(deterministically) create a list
of putative protein identifications. Specifically, for
each putative peptide identification it is straightforward to
determine, from a protein database, which
proteins contain that peptide. The information available can thus be
arranged in a hierarchical structure:
a list of $N$ putative protein identifications, with the information on
protein $k$ being
a list of $n_k$ putative peptide identifications, with a corresponding
vector of
scores $\mathbf{x}_k=(x_{k,1}, \ldots, x_{k, n_k})$. Here $x_{k,j}$ is
a scalar score that reflects how well
the spectrum associated with peptide $j$ in protein $k$ matches a
theoretical expectation under
the assumption that it was indeed generated by that peptide. (Typically
there are also
other pieces of information that are relevant in assessing the evidence
for peptide $j$ having
generated the spectrum, but we defer consideration of these to Section
\ref{SS:ancillary} below.)
In general, correct peptide identifications have higher scores than
incorrect ones, and proteins that are present tend to have more
high-scoring peptide identifications than the ones that are not
present. Our goal is to use this information to determine which
assembled proteins are present in the sample and which peptides are
correctly identified.

Note that, in the above formulation, if a peptide is contained in
multiple proteins, then the data for that
peptide is included multiple times. This is clearly sub-optimal,
particularly as we will treat the data on different proteins as
independent. The practical effect is that if one peptide has a very
high score,
and belongs to multiple proteins, then \textit{all} these proteins will
likely be identified as being present,
even though only one of them may actually be present. This
complication, where one peptide maps to multiple proteins, is referred
to as ``degeneracy'' [\citet{keller02}].
We refer to our current treatment of degeneracy as the ``nondegeneracy
assumption'' for the rest of the
text. We view extension of our method
to deal more thoroughly with degeneracy as an important area for future work.

We use indicators $T_k$ to represent whether a protein $k$ is present
($T_k=1$) or absent ($T_k=0$) in the sample, and indicators $P_{k,i}$
to represent whether a peptide $i$ on the protein $k$ is correctly
identified ($P_{k,i}=1$) or incorrectly identified ($P_{k,i}=0$). We
let $\pi_0^*$ and $\pi_1^*=1-\pi_0^*$ denote the proportions of absent
and present proteins respectively:
%
%e2.1 ###
\begin{equation}
\Pr(T_k = j) = \pi_j^* \qquad(k=1, \ldots, N; j=0,1).
\end{equation}
If a protein is absent,
we assume that all its constituent peptides must be incorrectly
identified; in contrast, if a protein
is present, then we allow that some of its constituent peptides may be
correctly identified,
and others incorrectly [Figure~\ref{F:flowchart}(c)]. Specifically, we
assume that given the protein indicators the peptide indicators are
independent and identically distributed, with
%
%e2.3 ###
%e2.2 ###
\begin{eqnarray}
\Pr(P_{k,i} = 0 \vert T_k=0) &=&1,
\\
\Pr(P_{k,i} = 0 \vert T_k=1) &=&\pi_1,
\end{eqnarray}
where $\pi_1$ denotes the proportion of incorrect peptides on
proteins that are present.

Given the peptide and protein indicators, we assume that the number of
peptides mapping to
a present (resp., absent) protein has distribution $h_1$
(resp., $h_0$), and that the
scores for correctly (resp., incorrectly) identified peptides
are independent draws from a distribution $f_1$ (resp., $f_0$).
Since present proteins will typically have more peptides mapping to
them, $h_1$ should be stochastically
larger than~$h_0$. Similarly, since
correctly-identified peptides will typically have higher scores, $f_1$
should be stochastically larger than $f_0$. The details on the choice
of functional form for these distributions
are discussed in Section \ref{SS:f} for $f_j$ and in Section \ref{SS:h}
for $h_j$.

Let $\Psi$ denote all the parameters in the above model, which include
$(\pi_0^*,\pi_1^*,\pi_1)$ as
well as any parameters in the distributions $h_0,h_1,f_0$ and $f_1$. We
will use $X,\mathbf{n}$ to denote the observed data, where $X=(\mathbf{x}_1,\dots,\mathbf{x}_N)$, and $\mathbf{n}=(n_1,\dots,n_N)$.
The above assumptions lead to the following nested mixture model:
%
%e2.4 ###
\begin{equation}\label{E:mixture0}
L(\Psi) = p(X,\mathbf{n}; \Psi) = \prod_{k=1}^{N}[\pi_0^*g_0(\mathbf
{x}_k)h_0(n_k)+\pi_1^*g_1(\mathbf{x}_k)h_1(n_k)],
\end{equation}
where
%
%e2.6 ###
%e2.5 ###
\begin{eqnarray}\label{E:g0}
g_0(\mathbf{x}_k) &\equiv& p(\mathbf{x}_k\vert n_k, T_k=0)=\prod
_{i=1}^{n_k} f_0(x_{k, i}),
\\\label{E:g1}
g_1(\mathbf{x}_k) &\equiv& p(\mathbf{x}_k\vert n_k, T_k=1)=\prod
_{i=1}^{n_k} [\pi_1 f_0(x_{k, i}) + (1-\pi_1) f_1(x_{k,i})].
\end{eqnarray}

Given the parameters $\Psi$,
the probability that protein $k$ is present in the sample can be
computed as
%
%e2.7 ###
\begin{equation} \label{eqn:protprob}
\Pr(T_k=j \vert\mathbf{x}_k, n_k; \Psi) = \frac{\pi_j^*g_j(\mathbf
{x}_k)h_j(n_k)}{\sum_{j=0,1}\pi_j^*g_j(\mathbf{x}_k)h_j(n_k) }.
\end{equation}

Similarly, the classification probabilities for peptides on the
proteins that are present are
%
%e2.8 ###
\begin{equation}
\Pr(P_{k,i}=1 \vert x_{k,i}, T_k=1; \Psi) =\frac{\pi_1 f_1(x_{k,i})}{\pi
_1 f_0(x_{k,i}) + (1-\pi_1) f_1(x_{k,i})}.
\end{equation}
As an absent protein only contains incorrect peptide identifications,
that is, $\Pr(P_{k,i}=1 \vert\mathbf{x}_k, T_k=0)=0$, the marginal
peptide probability is
%
%e2.9 ###
\begin{equation} \label{eqn:pepprob}
\Pr(P_{k,i}=1 \vert\mathbf{x}_k) = \Pr(P_{k,i}=1 \vert\mathbf{x}_k,
T_k=1)\Pr(T_k=1 \vert\mathbf{x}_k).
\end{equation}
This expression emphasizes how each peptide's classification
probability is affected by the classification probability of its parent protein.
We estimate values for these classification probabilities by estimating
the parameters $\Psi$
by maximizing the likelihood, (\ref{E:mixture0}),
and substituting these estimates into the above formulas.

The idea of modeling the scores of putative peptide identifications
using a mixture model
is also the basis of PeptideProphet [\citet{keller02}]. Our approach
here extends this
to a nested mixture model, modeling the overall sample as a mixture of present
and absent proteins. By simultaneously modeling the peptide and protein
classifications, we obtain
natural formulas, (\ref{eqn:protprob}) and (\ref{eqn:pepprob}), for the
probability that each protein
is present, and each peptide correctly identified.

It is helpful to contrast this approach with the
PeptideProphet/\break ProteinProphet two-stage strategy,
which we now describe in more detail.
First PeptideProphet models the overall sample as a mixture of present and
absent peptides, ignoring the information on which peptides map to
which proteins.
This leads naturally to a formula for the probability for each peptide
being correctly identified, $\Pr(P_{k,i}=1 | X)$, and these
probabilities are output by PeptideProphet. To translate these probabilities
into a measure of the strength of evidence that each \textit{protein} is
present, ProteinProphet
essentially uses the formula
%
%e2.10 ###
\begin{equation}
\Pr_{\mathrm{prod}}(T_k=1 | X)=1-\prod_{i} \Pr(P_{k,i}=0 | X),
\end{equation}
which we refer to as the ``product rule'' in the remainder of this text.
This formula is motivated by the idea that a protein should be called
as present only if not
all peptides mapping to it are incorrectly identified, and by treating
the incorrect identification of each peptide as independent (leading to
the product).

There are two problems with this approach.
The first is that the probabilities output by
PeptideProphet ignore relevant information on the nested structure
relating peptides
and proteins. Indeed, \citet{nesvizhskii03} recognize this problem, and
ProteinProphet actually makes an {ad hoc} adjustment to the
probabilities output by PeptideProphet, using the expected number of
other correctly-identified peptides on the same protein, before
applying the product rule. We will refer
to this procedure as the ``adjusted product rule.'' The second, more
fundamental, problem
is that the independence assumption underlying the product rule does
not hold
in practice. Indeed, there is a strong correlation among the
correct/incorrect statuses of peptides on the same protein.
For example, if a protein is absent, then (ignoring degeneracy) all its
constituent peptides must
be incorrectly identified. In contrast, our approach makes a very
different independence assumption,
which we view as more reasonable. Specifically, it assumes that, {\it
conditional on the correct/incorrect status of different peptides}, the
scores for different peptides
are independent.

Empirically, it seems that, despite these issues,
ProteinProphet is typically quite effective at identifying which proteins
are most likely to be present. However, as we show later, probabilities
output by
the product rule are not well calibrated, and there are settings in which
it can perform poorly.

%s2.3 ###
\subsection{Choice of scores and distributions $f_0$, $f_1$}\label{SS:f}

Recall that $f_0$ and $f_1$ denote the distribution of scores for
peptides that
are incorrectly and correctly identified. Appropriate choice of these
distributions may
depend on the method used to compute scores [\citet{choi08a}].
To facilitate comparisons
with PeptideProphet, we used the discriminant summary used by
PeptideProphet, \textit{fval}, as our score.
Of course, it is possible that other choices may give better performance.

Similar to ProteinProphet, when a single peptide is matched to multiple
spectra, each match
producing a different score, we summarized these data using the highest
score. (ProteinProphet keeps the one with the highest PeptideProphet
probability, which is usually, but not always, the one with the highest
score.) An alternative would be to
model all scores, and treat them as independent, as in [\citet{shen08}].
However, in preliminary empirical assessments we found using
the maximum produces better results, presumably because the
independence assumption is poor
(scores of spectra matching to the same peptide are usually highly
correlated [\citet{keller02}]).

%f2 ###
\begin{figure}[b]

\includegraphics{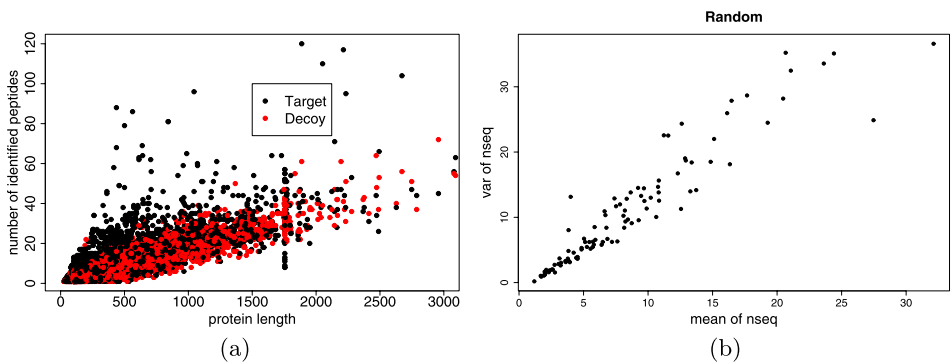}

\caption{Number of unique peptide hits and protein length in a yeast
data set. \textup{(a)} The relationship between number of peptide hits (Y-axis)
and protein length (X-axis). Red dots are decoy proteins, which
approximate absent proteins; black dots are target proteins, which
contain both present proteins and absent proteins. \textup{(b)} Verification of
the Poisson model for absent proteins, approximated by decoy proteins,
by mean--variance relationship. Proteins are binned by length with each
bin containing $1\%$ of data. Mean and variance of the number of
sequences are calculated for the observations in each bin.}\label{F:length}
\end{figure}

\begin{figure}[b]

\includegraphics{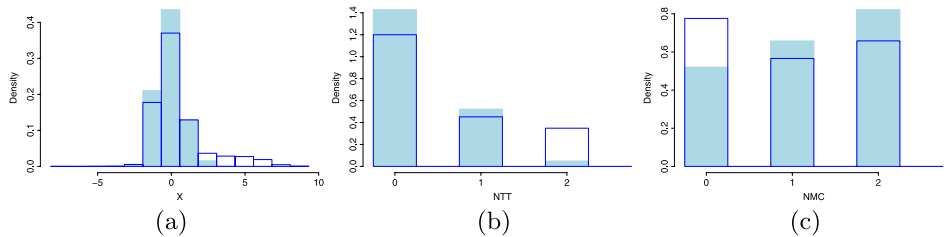}

\caption{The empirical distribution of features from peptide
identification in a yeast data set. Border histogram: real peptides,
which are a mixture of correct and incorrect identifications. Solid
histogram: decoy peptides, whose distribution approximates the
distribution of the incorrect identifications. \textup{(a)} Summary
score $X$. \textup{(b)} Number of tryptic termini. \textup{(c)} Number of
missing cleavages.}\label{F:dist}
\end{figure}

We chose to use a normal distribution, and shifted gamma distribution,
for $f_0$ and $f_1$:
\begin{eqnarray*}
f_0(x) &=& N(x;\mu, \sigma^2),
\\
f_1(x) &=& \operatorname{Gamma}(x;\alpha, \beta, \gamma),
\end{eqnarray*}
where $\mu$ and $\sigma^2$ are the mean and variance of the normal
distribution, and $\alpha$, $\beta$ and $\gamma$ are the shape
parameter, the scale parameter and the shift of the Gamma distribution.
These choices were made based on the shapes of the empirical
observations [Figure \ref{F:dist}(a)], the density ratio at the tails of
the distributions, and the goodness of fit between the distributions
and the data, for example, BIC [\citet{schwarz78}]. See \citet{li08b} for
further details. In particular, to assign peptide labels properly in
the mixture model, we require $f_0/f_1 > 1$ for the left tail of $f_0$,
and $f_1/f_0 > 1$ for the right tail of $f_1$.

Note that these distribution choices differ from PeptideProphet, which
models $f_0$ as shifted Gamma and $f_1$ as Normal. The distributions
chosen by PeptideProphet do not satisfy the requirement of $f_0/f_1$
above and can pathologically assign observations with low scores into
the component with higher mean.
The selected distributions fit our data well and also the data in Shen
et al., who chose the same distributions as ours after fitting a
two-component mixture model to the PeptideProphet discriminant summary
of their data. However, alternative distributions may be needed based
on the empirical data, which may depend on the choice of method for
assigning scores.

In this setting it is common to allow peptides with different charge
states to have different distributions of scores. This would be
straightforward, for example, by estimating the parameters of $f_0$ and
$f_1$ separately for different charge states. However,
in all the results reported here we do not distinguish charge states,
because in empirical comparisons we found that, once the ancillary
information in Section~\ref{SS:ancillary} was included, distinguishing
charge states made little difference to either the discriminating power
or the probability calibration. A similar result is reported in \citet{kall07}.

%s2.4 ###
\subsection{\normalsize Choice of $h$: incorporating protein
length}\label{SS:h}

Recall that $h_0$ and $h_1$ denote the distributions for $n_k$, the
number of putative identified
peptides on protein $k$, according to whether protein $k$ is absent or present.
It is known that long proteins tend to have more identified peptides
than short proteins (Figure \ref{F:length}), because of their potential
to generate more peptides in the experimental procedure, and the higher
chance to be randomly matched by incorrect peptide identifications. We
therefore allow the distribution of $n_k$ to depend on the protein
length $l_k$. Length correction, though of a different sort, has been
reported useful for reducing false identifications of long absent
proteins that are mapped by many incorrect identifications [\citet{price07}].

%f3 ###

It might be expected that the rate of incorrect peptide identification
in a fixed protein length is roughly uniform across all the proteins in
the database. Thus, we choose $h_0$ to be Poisson with mean $c_0l_k$,
where $c_0$ represents the average number of incorrect peptide
identifications in a unit protein length and is constant for all the
absent proteins. The mean--variance relationship of $n_k$ for absent
proteins in a real data set [Figure \ref{F:length}(b)] confirms that the
Poisson model is a reasonable fit.

For present proteins, we choose $h_1$ to be Poisson with mean $c_1l_k$,
where $c_1$ is a constant
that is bigger than $c_0$ to take account of the correct peptide
identifications additional to the incorrect ones. Similar Poisson
assumptions, though with different parameterization, were also made
elsewhere [\citet{price07}].

Because constructed proteins are assembled from one or more identified
peptides (i.e., $n_k > 0$), we truncate both Poisson distributions at
0, that is,
%
%e2.11 ###
\begin{equation}
h_j(n_k \vert l_k) = \frac{\exp(-c_j l_k) (c_j l_k)^{n_k}}{n_k! (1-\exp
(-c_j l_k))} \qquad (n_k = 1,2,\dots;  j=0,1).
\end{equation}

%s2.5 ###
\subsection{Incorporating ancillary information}\label{SS:ancillary}

In addition to the scores on each peptide identification based on the
spectra, other
aspects of identified peptide sequences, such as the number of tryptic
termini (NTT) and the number of missing cleavage (NMC), are informative
for the correctness of peptide identifications [\citet{kall07}; \citet{keller02};
\citet{choi08a}]. Because $\mathit{NTT} \in\{0,1,2\}$ [Figure \ref{F:dist}(b)], we model
it using a multinomial distribution. We discretize NMC, which usually
ranges from 0 to 10, into states (0, 1 and 2$+$) [Figure \ref{F:dist}(c)],
and also model it as a multinomial distribution. These treatments are
similar to PeptideProphet.

Peptide identification scores and features on peptide sequences have
been shown to be conditionally independent given the status of peptide
identification [\citet{keller02}; \citet{choi08a}]. Thus, we may incorporate the
ancillary information by replacing $f_j(X_{k,i})$ in (\ref{E:g0}) and
(\ref{E:g1}) with
$f_j(X_{k,i})f_j^{\mathit{NTT}}(\mathit{NTT}_{k,i})f_j^{\mathit{NMC}}(\mathit{NMC}_{k,i})$ ($j=0, 1$).
Further pieces of relevant information could be incorporated in a
similar way.

%s2.6 ###
\subsection{\normalsize Parameter estimation and initialization}\label{SS:est}

We use an expectation-maximi\-zation (EM) algorithm [\citet{DLR77}] to
estimate the parameters in our model and infer the statuses of peptides
and proteins, with the statuses of proteins ($T_k$) and peptides
($P_{k,i}$) as latent variables. The augmented data for protein $k$
take the form of $Y_k \equiv(\mathbf{X}_k, n_k, T_k,\break P_{k,1}, \ldots,
P_{k, n_k})$. The details of the EM algorithm can be found in the \hyperref[app]{Appendix}.

To select a reasonable starting point for the EM algorithm, in the real
data set, we initialize the parameters related to incorrect peptide
identification ($f_0, f_0^{\mathit{NTT}}, f_0^{\mathit{NMC}}$ and $c_0$) using estimates
obtained from the decoy database (see Section \ref{S:yeast} for
details). For $f_1$, we initialize the shift $\gamma^{(0)}=\min_{k,
i}(x_{k,i})-\varepsilon$, where~$\varepsilon$ is a small positive number to
ensure $x_{k,i}-\gamma^{(0)} > 0$ for all identified peptides (in both
real and decoy databases), and estimate $\alpha$ and $\beta$ using the
sample mean and sample variance of the scores. We initialize
$f_1^{\mathit{NTT}}$ and $f_1^{\mathit{NMC}}$ using the peptides that are identified in
the real database and are scored in the upper $90\%$ of the
identifications to the real database. As $c_1 > c_0$, we choose $c_1 =
b c_0$, where $b$ is a random number in $[1.5, 3]$. The starting values
of $\pi_0^*$ and $\pi_1$ are chosen randomly from $(0,1)$. For each
inference, we run
the EM algorithm from 10 random starting points and report the results
from the run converging to the highest likelihood.

%s3 ###
\section{Results} \label{S:Results}

%s3.1 ###
\subsection{Simulation studies}\label{S:simulation}

We first use simulation studies to examine the performance of
our approach, and particularly to assess the potential for it
to improve on the types of
2-stage approach used by PeptideProphet and ProteinProphet.
Our simulations are generated based on our nested mixture model, and
ignore many of the complications of real data (e.g., degeneracy). Thus,
their primary goal is not to provide evidence that our approach is
actually superior in practice. Rather, the
aim is to provide insight into the kind of gains in performance that
\textit{might} be achievable in practice, to illustrate settings where the
product rule used by ProteinProphet may perform
particularly badly, and to check for robustness of our method to one of
its underlying assumptions (specifically the assumption that the
expected proportion of incorrect peptides is the same for all present proteins).
In addition, they provide a helpful check on the correctness of our EM
algorithm implementation.

At the peptide level, we compare results from our model with the
peptide probabilities computed by PeptideProphet, and the
PeptideProphet probabilities adjusted by ProteinProphet (see Section
\ref{SS:nested}). At the protein level, we compare results from our
model with three methods: the classical deterministic rule that calls a
protein present if it has two or more high-scored peptides (which we
call the ``two-peptide rule'') and the two product rules (adjusted and
unadjusted; see Section \ref{SS:nested}). Because the product rule is
the basis of ProteinProphet, the comparison with the product rule
focuses attention on the fundamental differences between our method and
ProteinProphet, rather than on the complications of degeneracy handling
and other heuristic adjustments that are made by the ProteinProphet software.

%t1 ###
\begin{table}
\tabcolsep=0pt
\caption{Simulation parameters and parameter estimation in the
simulation studies. The simulation parameters are estimated from a
yeast data set. $\pi_0$ is the proportion of incorrect peptides on the
absent proteins in the simulated data}\label{T:para-est1}
\fontsize{8}{8}\selectfont{\begin{tabular*}{\tablewidth}{@{\extracolsep{4in minus 4in}}lcccccccc@{}}
\hline
& & $\bolds{\pi_0^*}$ & $\bolds{c_0}$ & $\bolds{c_1}$ & $\bolds{\pi_0}$ & $\bolds{\pi_1}$ & $\bolds{f_0}$ & $\bolds{f_1}$ \\
\hline
S1 &True parameter & 0.88 & 0.018 & 0.033 & 1\phantom{.000}& 0.58 & $G(86.46, 0.093,
-8.18)$ & $N(3.63, 2.07^2)$
\\
& Estimated values & 0.87 & 0.018 & 0.032 & -- & 0.58 & $G(86.24, 0.093,
-8.18)$ & $N(3.57, 2.05^2)$
\\[3pt]
S2 & True parameter & 0.5 & 0.018 & 0.033 & 0.998& 0.58 & $G(86.46,
0.093, -8.18)$ & $N(3.63, 2.07^2)$
\\
& Estimated values & 0.55 & 0.018 & 0.034 & -- & 0.56 & $G(83.78, 0.096,
-8.18)$ & $N(3.71, 2.08^2)$
\\[3pt]
S3 &True parameter & 0.88 & 0.018 & 0.033 & 1\phantom{.000}& $\operatorname{Unif}(0, 0.8)$ &
$G(86.46, 0.093, -8.18)$ & $N(3.63, 2.07^2)$
\\
& Estimated values & 0.88 & 0.018 & 0.034 & -- & 0.40 & $G(85.74, 0.094,
-8.18)$ & $N(3.68, 2.05^2)$
\\ \hline
\end{tabular*}}
\end{table}

As PeptideProphet uses Gamma for $f_0$ and Normal for $f_1$, we follow
this practice in the simulations (both for simulating the data and
fitting the model). In an attempt to generate realistic simulations, we
first estimated parameters from a yeast data set  [\citet{kall07}] using
the model in Section \ref{S:methods}, except for this change of $f_0$
and $f_1$, then simulated proteins from the estimated parameters (Table
\ref{T:para-est1}).

We performed three simulations, S1, S2 and S3, as follows:
\begin{enumerate}[S1.]
\item[S1.] This simulation was designed to demonstrate performance when
the data are generated from the same nested mixture model we use for
estimation. Data were simulated from the mixture model,
using the parameters estimated from the real yeast data set considered
below. The resulting data contained 12\% present proteins and 88\%
absent proteins, where protein length $l_k \sim\exp(1/500)$.
\item[S2.] Here simulation parameters were chosen to illustrate a
scenario where the product rule performs particularly poorly. Data were
simulated as in S1, except for (i) the proportion of present proteins
was increased to 50\% ($\pi_0^*=0.5$); (ii) the distribution of protein
length was modified so that all present proteins were short ($l_k \in
[100,200]$) and absent proteins were long ($l_k \in[1000, 2000]$); and
(iii) we allowed that absent proteins may have
occasional high-scoring incorrect peptide identifications ($0.2\%$ of
peptide scores on absent proteins were drawn from $f_1$ instead of $f_0$).
\item[S3.] A simulation to assess sensitivity of our method to
deviations from the assumption that the proportion of incorrect
peptides is the same for all present proteins. Data were simulated as
for S1, except $\pi_1 \sim \operatorname{Unif}(0, 0.8)$ independently for each present protein.
\end{enumerate}

In each simulation, 2000 proteins were simulated. We forced all
present proteins to have at least one correctly identified peptide. For
simplicity, only one identification score was simulated for each
peptide, and the ancillary features for all the peptides ($\mathit{NMC}=0$ and
$\mathit{NTT}=2$) were set identical. We ran the EM procedure from several random
initializations close to the simulation parameters. We deemed
convergence to be achieved when the log-likelihood increased $<$0.001
in an iteration. PeptideProphet (TPP version 3.2) and ProteinProphet
(TPP version 3.2) were run using their default values.

\subsubsection*{Parameter estimation}
In all the simulations, the parameters estimated from our models are
close to the true parameters (Table \ref{T:para-est1}). Even when
absent proteins contain a small proportion of high-scored peptides (S2)
or the assumption of a fixed $\pi_1$ is violated (S3), our method still
produces reasonable parameter estimations.

\subsubsection*{Tradeoff between true calls and false calls}
We compared the performances of different methods by the tradeoff
between the number of correct and incorrect calls made at various
probability thresholds. As a small number of false calls is desired in
practice, the comparison focuses on the performance in this region.

\begin{figure}

\includegraphics{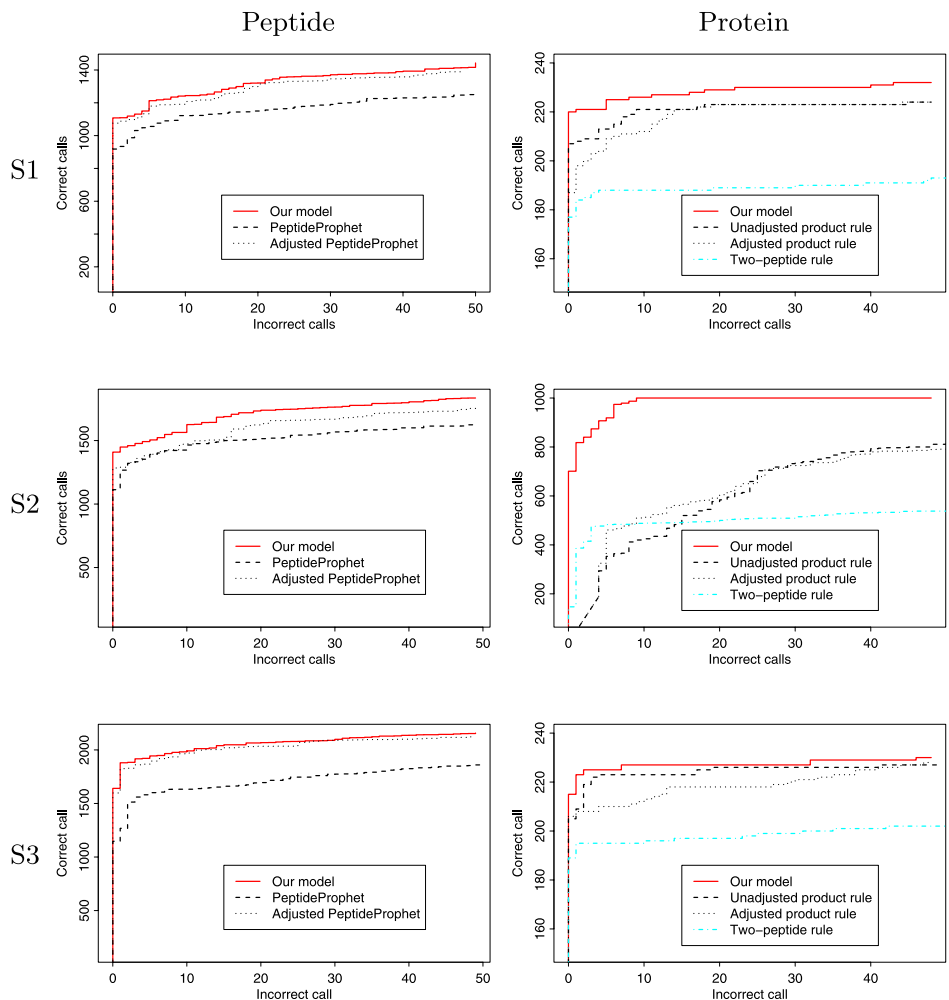}

\caption{The number of correct and incorrect calls made at various
thresholds in simulation studies. Incorrect calls: the number of
incorrect peptides or absent proteins assigned posterior probabilities
exceeding the thresholds; correct calls: the number of correct peptides
or present proteins assigned posterior probabilities exceeding the
thresholds.}\label{F:IT-IP1}
\end{figure}

At the peptide level, our model consistently identifies substantially
more ($>$100 in all cases) true peptides than PeptideProphet at any
controlled number of false peptides in the range of 0--200 [Figure \ref
{F:IT-IP1} (S1) left and (S2) left], in all the simulations. This gain illustrates the
potential for our one-stage model to provide effective feedback of
information from the protein level to peptide level, to improve peptide
identification accuracy.

At the protein level, our model consistently identifies more true
proteins than the adjusted product rule at any controlled number of
false proteins in the range of 0--50, in all simulations [Figure \ref
{F:IT-IP1} (S1) right and (S2) right]. In S2 the product rules perform less well than the
other two simulations. This poor performance is anticipated in this
setting, due to its assumption that correctness of peptides on the same
proteins is independent. In particular, when absent proteins with big
$n_k$ contain a single high-scored incorrect peptide, the product rule
tends to call them present. When present proteins with small $n_k$
contain one or two correct peptides with mediocre scores besides
incorrect ones, the product rule tends to call them absent. The
examination of individual cases confirms that most mistakes made by the
product rule belong to either of the two cases above.

It is interesting that although the adjusted product rule improves
peptide identification accuracy compared with the unadjusted rule, it
also worsens the accuracy of protein identification (at least in S1 and
S3). This illustrates a common pitfall of {\em ad hoc} approaches:
fixing one problem may unintentionally introduce others.

%f4 ###

\subsubsection*{Calibration of probabilities}
\begin{figure}

\includegraphics{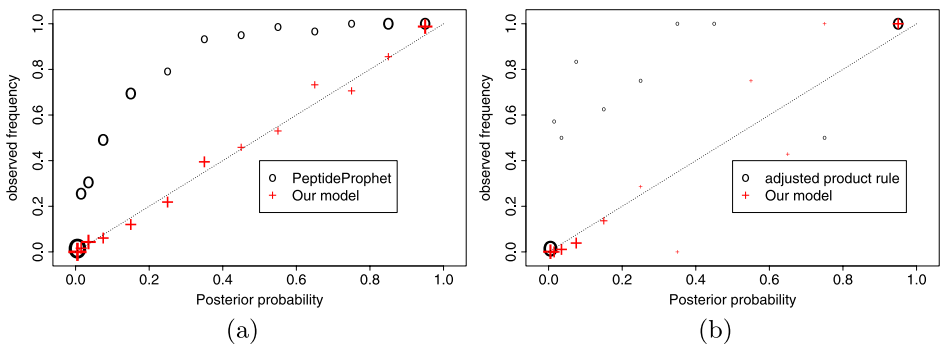}

\caption{Calibration of posterior probabilities in a simulation study
\textup{S1}. The observations are binned by the assigned probabilities. For
each bin, the assigned probabilities (X-axis) are compared with the
proportion of identifications that are actually correct (Y-axis). \textup{(a)}
peptide probabilities, \textup{(b)} protein probabilities. Black:
PeptideProphet [in \textup{(a)}] or adjusted product rule [in \textup{(b)}]; Red: our
method. The size of the points represents the number of observations in
each bin. Other simulations have similar results.}\label{F:calibrate1}
\end{figure}

Methods for identifying proteins and peptides should, ideally, produce
approximately calibrated probabilities, so that the estimated posterior
probabilities can be used as a way to assess the uncertainty of the
identifications. In all the three simulations the peptide probabilities
from our method are reasonably well calibrated, whereas the
PeptideProphet probabilities are not, being substantially smaller than
the actual probabilities [Figure \ref{F:calibrate1}(a)]. Our method
seems to be better calibrated than the adjusted product rule at the
protein level [Figure \ref{F:calibrate1}(b)]. However, very few proteins
are assigned probabilities${}\in[0.2,0.9]$, so larger samples would be
needed to confirm this.

%f5 ###

%s3.2 ###
\subsection{A standard mixture}

Mixtures of standard proteins have been used for assessing the
performance of identifications. Although these mixtures are known to be
too simple to reflect the complexity of the realistic samples and may
contain many unknown impurities [\citet{elias05}], they can nonetheless
be helpful as a way to assess whether a method can effectively identify
the known components.

We applied our method on a standard protein mixture [\citet{purvine04}]
used in \citet{shen08}. This data set consists of the MS/MS spectra
generated from a sample composed of 23 stand-alone peptides and trypsin
digest of 12 proteins. It contains three replicates with a total of
9057 spectra. The experimental procedures are described in \citet
{purvine04}. We used Sequest [\citet{eng94}] to search, with
nontryptic peptides allowed, a database composed of the 35
peptides/proteins, typical sample contaminants and the proteins from
\textit{Shewanella oneidensis}, which are known to be not present in the
sample and serve as negative controls. After matching spectra to
peptides, we obtained 7935 unique putative peptide identifications. We
applied our methods to these putative peptide identifications, and
compared results, at both the protein and peptide levels, with results
from the same standard mixture reported by Shen et al. for both their
own method (``Hierarchical Statistical Method''; HSM) and for
PeptideProphet/ProteinProphet. Note that in assessing each method's
performance we make the assumption, standard in this context, that a
protein identification is correct if and only if it involves a known
component of the standard mixture, and a peptide identification is
correct if and only if it involves a peptide whose sequence is a
subsequence of a constituent protein (or is one of the 23 stand-alone peptides).

\begin{figure}

\includegraphics{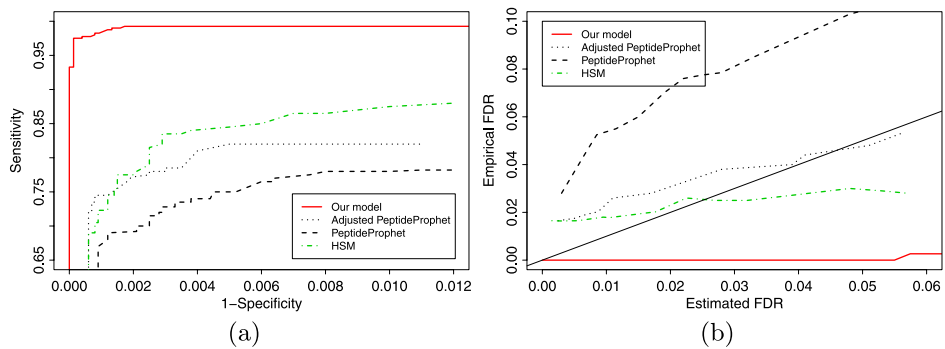}

\caption{Peptide identification on a standard protein mixture. \textup{(a)} ROC
curves for peptide identification on a standard protein/peptide
mixture. \textup{(b)} Calibration of the FDR estimates for peptide
identifications on a standard protein/peptide mixture. The straight
line represents a perfect estimate.}\label{F:12mixture}
\end{figure}

At the protein level all of the methods we compare here identify all 12
proteins with probabilities close to 1 before identifying any false
proteins. Our method provides a bigger separation between the
constituent proteins and the false proteins, with the highest
probability assigned to a false protein as 0.013 for our method and
above 0.8 for ProteinProphet and HSM.
At the peptide level, our model shows better discriminating power than
all the other methods [Figure \ref{F:12mixture}(a)]. Again, we ascribe
this better performance at the peptide level to the ability of our
model to effectively feedback information from the protein level to the
peptide level.\looseness=1

To assess calibration of the different methods for peptide
identification, we compare the empirical FDR and the estimated FDR
[Figure \ref{F:12mixture}(a)], where the estimated FDR is computed as the
average posterior probabilities to be absent from the sample for the
identifications [\citet{efron01}; \citet{newton04}]. None of the methods is
particularly well-calibrated for these data: our method is conservative
in its
estimated FDR, whereas the other methods tend to underestimate FDR at
low FDRs. Our conservative estimate of FDR in this case partly reflects
the simplicity of this artificial problem. Indeed, our method
effectively separates out the real and not real peptides almost
perfectly in this case: 99\% of peptide identifications are assigned
probability either $>$0.99 (all of which are real) or $<$0.01 (less
than one percent of which are real). Thus, for both these groups our
method is effectively calibrated. The conservative calibration apparent
in Figure \ref{F:12mixture}(b) reflects the fact that the remaining 1\%
of peptides that are assigned intermediate probabilities (between 0.01
and 0.99) are all real.

%f6 ###

We emphasise that this standard mixture inevitably provides only a very
limited comparison of the performance of different methods. Indeed, the
fact that in this case all methods effectively correctly identify all
the real proteins, with no false positives, suggests that this standard
mixture, unsurprisingly, provides nothing like the complexity of most
real data problems. On the other hand, it is reassuring to see our
method perform well on this problem, and the results in Figure \ref
{F:12mixture}(a) do provide a nice illustration of the potential benefits
of effective feedback of information from the protein to peptide level.

%s3.3 ###
\subsection{Application on a yeast data set}\label{S:yeast}

To provide comparisons on more realistic data, we also compared methods
using a yeast data set [\citet{kall07}]. Because the true protein
composition of this data set is unknown, the comparisons were done by
use of a decoy database of artificial proteins,
which is a commonly used device in this setting [\citet{elias07}].
Specifically, in the initial step of matching spectra to peptides, each
spectrum was searched against a combined database, containing both
target (i.e., real) proteins and decoy (i.e., nonexistent) proteins
created by permuting the sequences in the target database. This search
was done using Sequest [\citet{eng94}]. The methods
are then applied to the results of this search, and they assign
probabilities to both target and
decoy proteins. Since the decoy proteins cannot be present in the
sample, and assuming
that their statistical behavior is similar to real proteins that are
absent from the sample, a false discovery
rate for any given probability threshold can be estimated by counting
the number of decoy
proteins assigned a probability exceeding the threshold.

The data set contains 140,366 spectra. After matching spectra to
peptides (using Sequest [\citet{eng94}]), we obtained 116,264 unique
putative peptide identifications. We used DTASelect [\citet{tabb02}] to
map these peptides back to 12,602 distinct proteins (the proteins were
found using DTASelect [\citet{tabb02}]).

We compared our algorithm with PeptideProphet for peptide inferences
and actual ProteinProphet for protein inferences on this data set. The
HSM method, whose computational cost and memory requirement are
proportional to the factorial of the maximum protein group size,
encountered computation difficulties on this data set and failed to
run, because this data set contains several large protein groups. We
initialized our algorithm using the approach described in Section \ref
{SS:est}, and stopped the EM algorithm when the change of
log-likelihood is smaller than 0.001. PeptideProphet and ProteinProphet
were run with their default settings.

In this case the comparison is complicated by the presence of peptides
belonging to multiple proteins, that is, degeneracy, which occurs in
about $10\%$ of proteins in yeast. Unlike our approach, ProteinProphet
has routines to handle degenerate peptides. In brief, it shares each
such peptide among all its corresponding proteins, and estimates an
{ad hoc} weight that each degenerate peptide contributes to each
protein parent.
In reporting results, it groups together proteins with many shared
peptide identifications, such as homologs, and reports a probability
for each group (as one minus the product of the
probabilities assigned to each of the individual proteins being
absent). In practice, this has
the effect of upweighting the probabilities assigned to large groups
containing many proteins.

To make our comparison, we first applied our model ignoring the
degeneracy issue to compute
a probability for each protein being present, and then used these to
assign a probability to each
group defined by ProteinProphet. We treated proteins that were not in a
group as a group
containing one protein. For our method, we assigned to each group the
maximum probability assigned to any protein in the group. This also has
a tendency to upweight probabilities to large groups, but not
by as much as the ProteinProphet calculation.

%f7 ###
\begin{figure}[b]

\includegraphics{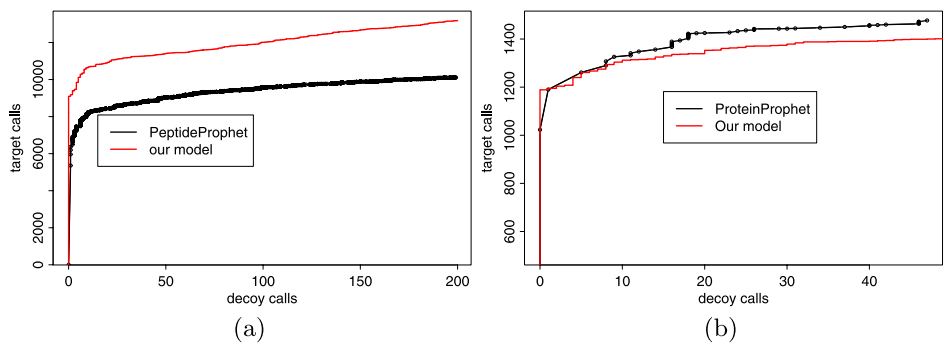}

\caption{The number of decoy and target peptides \textup{(a)} or protein groups
\textup{(b)} assigned probabilities exceeding various thresholds in a yeast data
set. Decoy calls: the number of decoy peptides or protein groups
assigned a probability exceeding the threshold. Target calls: the
number of target peptides or protein groups assigned a probability
exceeding the threshold. \textup{(a)} Peptide inference. \textup{(b)} Protein inference.}\label{F:tradeoff}
\end{figure}

Note that the tendency of both methods to upweight probabilities
assigned to large groups, although a reasonable thing to do, makes
reliably estimating the FDR more difficult. This is because, unlike the
real proteins, the decoy proteins do not fall into homologous groups
(i.e., each is in a group by itself), and so the statistical behavior
of the decoy groups will not exactly match those of the absent real
protein groups. The net effect will be that, for both methods, the
estimates of the FDR based on the decoy comparison will likely
underestimate the true FDR. Further, we suspect that the amount of
underestimation of the FDR will be stronger for ProteinProphet than for
our method, because ProteinProphet more strongly upweights
probabilities assigned to large groups. As a result, comparing the {\it
estimated} FDRs from each method, as we do here, may give a slight
unfair advantage to ProteinProphet. In any case, this rather subtle
issue illustrates the severe challenges of reliably comparing different
approaches to this problem.

We assessed the methods by comparing the number of target and decoy
protein groups assigned
probabilities exceeding various thresholds. We also compared the number
of decoy and target
peptides assigned probabilities exceeding various thresholds. The
results are shown in Figure \ref{F:tradeoff}.

At a given number of decoy peptide identifications, our model
identified substantially more target peptides than PeptideProphet
[Figure \ref{F:tradeoff}(a)]. Among these, our method identified
most of the target peptides identified by PeptideProphet, in addition
to many more not identified by PeptideProphet. For example, at $\mathit{FDR}=0$
(i.e., no decoy peptides identified), our method identified 5362
peptides out of 5394 peptides that PeptideProphet identified, and an
additional 3709 peptides that PeptideProphet did not identify.

For the protein idenfication, the methods identified similar numbers of
real protein groups at
small FDRs ($<$10 decoy proteins identified). At slightly larger FDRs
($>$10 decoy proteins identified)
ProteinProphet identified more real protein groups ($< $100) than our
method. This apparent slightly superior performance of ProteinProphet
may be due, at least in part, to issues noted above regarding likely
underestimation of the FDR in these experiments.

%s4 ###
\section{Comparison with HSM on another yeast data set}\label{S:yeast2}

To provide comparisons with HSM method on a realistic data set, we
compared our method with HSM on another yeast data set, which was
original published in \citet{elias05} and analyzed by  \citet
{shen08}. We were unable to obtain the data from the original
publication; instead, we obtained a processed version from Shen, which
produces the results in \citet{shen08}. Because the
processed data lacks several key features for processing by
PeptideProphet and ProteinProphet, we were unable to compare with
PeptideProphet and ProteinProphet on this data set.

This data set was generated by searching a yeast sample against a
sequence database composed of 6473 entries of yeast (\textit{Saccharomyces
cerevisiae}) and 22,437 entries of \textit{C. elegans} (\textit{Caenorhabditis
elegans}). In total, 9272 MS/MS spectra were assigned to 4148 unique
peptides. Following \citet{shen08}, we exclude 13 charge
$+$1 peptides and fit peptides with different charge states separately
(charge~$+$2: 6869 and charge $+$3: 2363). The rest of the 4135 peptides
consists of 3516 yeast peptides and 696 \textit{C. elegans} peptides. These
peptides map to 1011 yeast proteins and 876 \textit{C. elegans} proteins. Among
all the peptides, 468 (11.3\%) are shared by more than one protein and
77 peptides are in common between the two species. Due to peptide
sharing between species, 163 \textit{C. elegans} proteins contain only peptides
that are in common with yeast proteins. These proteins and peptides
shared between species are removed at performance evaluation for all
methods of comparison.

\begin{figure}

\includegraphics{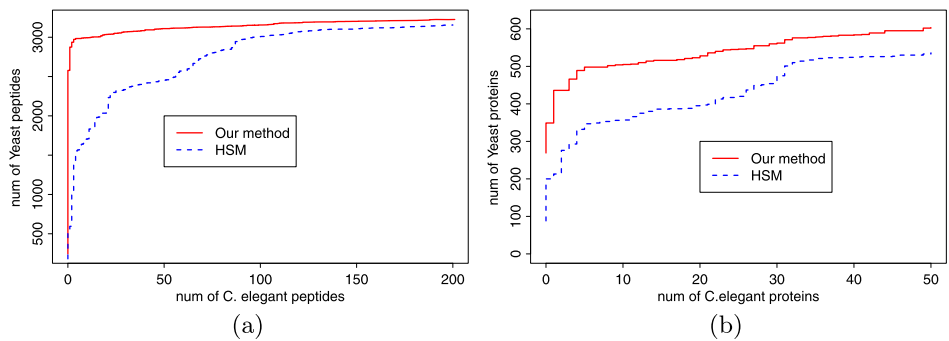}

\caption{The number of C. elegans and Yeast peptides \textup{(a)} or proteins
\textup{(b)} assigned probabilities exceeding various thresholds in the yeast
sample in Shen's paper.}\label{F:shen-data2}
\end{figure}

We compare the performance of our method with Shen's method for both
peptide inferences and protein inferences in Figure \ref{F:shen-data2}.
Similar to the previous section, a~false discovery rate for any given
probability threshold can be estimated by counting the number of \textit{C.
elegans} proteins assigned a probability exceeding the threshold, since
the \textit{C. elegans} peptides or proteins that do not share common sequences
with Yeast peptides or proteins cannot be present in the sample. We
assessed the methods by comparing the number of yeast and \textit{C. elegans}
peptides or proteins assigned probabilities exceeding various
thresholds. The results are shown in Figure~\ref{F:shen-data2}.

%f8 ###

At a given number of \textit{C. elegans} peptide identifications, our model
identified substantially more yeast peptide identifications than HSM at
small FDR ($< $100 \textit{C. elegans} peptides). For example, at $\mathit{FDR}=0$, our
method identifies 516 peptides out of 522 peptides that are identified
by HSM and an additional 2116 peptides that HSM did not identify. The
methods identified similar numbers of yeast peptides at higher FDR ($>
$100 \textit{C. elegans} peptides). For the protein identification, in the range
of FDR that we studied, our method consistently identifies over 80 more
yeast proteins than HSM at a given number of \textit{C. elegans} protein
identifications, in addition to the majority (e.g., 96.5\% at $\mathit{FDR}=0$) of
the yeast proteins identified by HSM.

Although ProteinProphet results reported by Shen et al. (Table 1 in
Shen et~al.) appear to identify more yeast proteins than our method at
a given number of \textit{C. elegans} proteins in the range they studied,
without access to the raw data, it is difficult to gain insights into
the differences. For example, the information on whether the reported
ProteinProphet identifications are proteins or protein groups and which
proteins are grouped together by ProteinProphet are unavailable from
the data we worked on. However, they are critical for making
comparisons on the same basis. The comparison with proper handling of
these issues (e.g., grouping our protein identifications as in Section
\ref{S:yeast}) may lead to conclusions different from a naive comparison.

%s5 ###
\section{Discussion}\label{S:discussion}

We have presented a new statistical method for assessing evidence for
presence of proteins and constituent peptides identified from mass
\mbox{spectra}. Our approach is, in essence, a model-based clustering method
that simultaneously identifies which proteins are present, and which
peptides are correctly identified. We illustrated the potential for
this approach to improve accuracy of protein and peptide identification in
both simulated and real data.

A key feature of our nested mixture model is its ability to incorporate
evidence feedback from proteins to the peptides nested on them. This
evidence feedback helps distinguish peptides that are correctly
identified but with weak scores, from those that are incorrectly
identified but with higher scores. The use of a coherent statistical
framework also avoids problems with what we have called the ``product
rule,'' which is adopted in several protein identification approaches
 [\citet{nesvizhskii03}; \citet{price07}], but is based on an inappropriate
assumption of independence of the presence and absence of
different peptides. It has been noted [e.g., \citet{sadygov04}; \citet{feng07}]
that the product rule tends to wrongly identify as present long
proteins with occasional high-scored incorrect peptides; our simulation
results [Figure \ref{F:IT-IP1}(S2)] illustrate this problem, and
demonstrate that our approach does not misbehave in this
way.\looseness=1

In recent work \citet{shen08} also introduced a nested
latent-variable-based model (HSM) for jointly identifying peptides and
proteins from MS/MS data. However, although HSM shares with our model
the goal of simultaneous modeling of peptides and proteins, the
structure of their model is different, and their approach also differs
in several details. Among these differences, the
following seem to us most important:
\begin{enumerate}[1.]
\item HSM accounts for degeneracy, whereas ours does not. We comment
further on this below.
\item HSM includes all the scores for those peptides that match more
than one spectrum, whereas our model uses only the maximum score as a
summary of the evidence.
Modeling all scores is obviously preferable in principle, but, in
practice, it is possible
that it could actually decrease identification accuracy. We note two
particular issues here:
(a) Shen et al. assume that, conditional on a peptide's
presence/absence status, multiple scores for the same peptide are
independent. This independence assumption will not hold in practice,
and the costs of such modeling errors could outweigh the benefits of
using multiple scores; (b) HSM appears to condition on the number of
spectra matched to each peptide, rather than treating this number as an
informative piece of data. As a result of this conditioning,
additional low-scoring hits to a peptide will always decrease the
probability assigned to that peptide.
This contrasts with our intuition that additional hits to a peptide
could, in some cases, increase confidence that it is present, even if
these hits have low scores.
\item HSM incorporates only whether the number of hits to peptides in a
protein exceeds some threshold, $h$ (which is set to 1 in their
applications). In contrast, our model incorporates the actual number of
(distinct) peptides hitting a protein using a Poisson model.
In this way our model
uses more available information, and accounts for variations in protein length.
Note that modeling only whether the number of hits exceeds $h$ has some
undesirable consequences, similar to those noted above for conditioning
on the number of hits to a peptide. For example, if $h=1$, then a
protein that has two hits, each with low scores, will be assigned a higher
identification probability than a protein that is hit more than twice
with low scores.
\item HSM conditions on the number of specific cleavages (NTT in our
development here)
in each putative peptide. Specifically, their parameter $\pi_{ij}(\alpha
)$ is
the probability of a particular cleavage event occurring, conditional
on NTT. In contrast, our model treats the NTT for each peptide hit as
observed data. This may improve identification accuracy because the
distribution of NTT differs greatly for correct and incorrect
identifications (Figure \ref{F:dist}).
\end{enumerate}

We expect that some of these differences in detail, perhaps in addition
to other differences not noted here, explain the different performances
of our method and that of Shen et al. on the standard mixture data and
the yeast data used in \citet{shen08}.
On the other hand, we agree with Shen et al. that comparisons like
these are less definitive, and harder to interpret, than one would
like, because of the absence of good gold-standard realistic data sets
where the truth is known.\looseness=1

We emphasize that, despite its promise, we view the model we present
here as only a starting point toward the development of more
accurate protein and peptide identification software. Not only is the
development of robust fast user-friendly software a considerable task
in itself, but there are also important aspects of real data---specifically degeneracy, which is
prevalent in high-level organisms---that are not properly accounted
for by our model.
Currently, most existing approaches to handle degeneracy are based on
heuristics. For example, ProteinProphet groups the proteins with shared
peptides and assigns weights to degenerate peptides using heuristics.
An exception is Shen et al.'s model [\citet{shen08}], which attempts to
provide a coherent statistical solution to the problem by
allowing that a peptide will be present in the digested sample if any
one of the
proteins containing that peptide generates it, and assuming that these
generation events are independent [their equation (2)]. However,
because their model computes all the possible combinations of protein
parents, which increases in the order of factorials, it is
computationally prohibitive to apply their method on data with a
moderate or high degree of degeneracy.
It should be possible to extend our model to allow for degeneracy in a
similar way. However, there are some steps that
may not be straightforward. For example, we noted above that our model
uses NTT as observed data. But under degeneracy, NTT for each peptide
is not directly observed, because
it depends on which protein generated each peptide. Similarly, the
number of distinct peptides identified on each protein depends
on which protein generated each peptide. While it should be possible to
solve these issues by introducing appropriate latent variables,
some care may be necessary to ensure that, when degeneracy is accounted
for, identification accuracy improves as it should.

\begin{appendix}
\section*{Appendix}\label{app}

Here we describe an EM algorithm for the estimation of $\Psi=(\pi_0^*,
\pi_0, \pi_1, \mu, \sigma,\break \alpha, \beta, \gamma, c_0, c_1)^T$ and the
protein statuses and the peptide statuses. To proceed, we use $T_k$ and
$(P_{k, 1}, \ldots, P_{k, n_k})$ as latent variables, then the
complete
log-likelihood for the augmented data $Y_k \equiv$ $(\mathbf{X}_k, n_k,
T_k,$ $P_{k, 1}, \ldots,$ $P_{k, n_k})$ is
%
%e5.1 ###
\begin{eqnarray}
&&l^C(\Psi\vert\mathbf{Y})\nonumber
 \\
&&\qquad =\sum_{k=1}^N \Biggl\{(1-T_k)\Biggl[\log\pi_0^* + \log h_0(n_k \vert l_k, n_k >0)\nonumber
\\
&&{}\hspace*{62pt}\qquad\quad +\sum_{i=1}^{n_k}(1-P_{k,i}) \log(\pi_0 f_0(x_{k,i}))\nonumber
\\
&&{}\hspace*{62pt}\qquad\quad+\sum_{i=1}^{n_k} P_{k,i} \log\bigl((1-\pi_0)f_1(x_{k,i})\bigr)\Biggr]\Biggr\}
\\
&&{}\qquad\quad + \sum_{k=1}^N \Biggl\{T_k\Biggl[\log(1-\pi_0^*) + \log h_1(n_k \vert l_k, n_k
>0)\nonumber
\\
&&{}\hspace*{50pt}\qquad\quad+\sum_{i=1}^{n_k}(1-P_{k,i}) \log(\pi_1 f_0(x_{k,i}))\nonumber
\\
 &&{}\hspace*{64pt}\qquad\quad+\sum_{i=1}^{n_k} P_{k,i} \log\bigl((1-\pi_1)f_1(x_{k,i})\bigr)\Biggr]\Biggr\}.\nonumber
\end{eqnarray}

E-step:
%
%e5.2 ###
\begin{eqnarray}
Q\bigl(\Psi, \Psi^{(t)}\bigr) &\equiv& E\bigl(l^C(\Psi) \vert\mathbf{x}, \Psi^{(t)}\bigr)\nonumber
\\[2pt]
\nonumber
 &=& \sum_{k=1}^{N} P(T_k=0)\Biggl\{\log{\pi_0^*} + \log{h_0(n_k\vert l_k, n_k > 0)}
\\ \nonumber
&&\hspace*{66pt}{} + \sum_{i=1}^{n_k}P(P_{k,i}=0 \vert T_k=0)\log(\pi_0f_0(x_{k,i}))
\\[2pt]
&&\hspace*{66pt}{} +
\sum_{i=1}^{n_k}P(P_{k,i}=1 \vert T_k=0)\log\bigl((1-\pi_0)f_1(x_{k,i})\bigr)\Biggr\}\mathrm{}
 \\[2pt]
\nonumber
&&{}\hspace*{-2pt}+ \sum_{k=1}^{N} P(T_k=1)\Biggl\{\log(1-\pi_0^*) + \log{h_1(n_k\vert l_k, n_k> 0)}
\\ \nonumber
&&\hspace*{78pt}{}+ \sum_{i=1}^{n_k}P(P_{k,i}=0 \vert T_k=1)\log(\pi_1f_0(x_{k,i}))
 \\[2pt]
 &&\hspace*{78pt}{} +
\sum_{i=1}^{n_k}P(P_{k,i}=1 \vert T_k=1)\log\bigl((1-\pi_1)f_1(x_{k,i})\bigr)\Biggr\}.\nonumber
\end{eqnarray}

Then
%
%e5.3 ###
\begin{eqnarray}\label{3E:IT}
\hat{T}_k^{(t)} &\equiv& E\bigl(T_k \vert\mathbf{x}_k, n_k, \Psi^{(t)}\bigr)\nonumber
 \\[-1pt]
&=& \frac{P(T_k=1, \mathbf{x}_k, n_k \vert\Psi^{(t)})}{P(\mathbf{x}_k,
n_k \vert\Psi^{(t)})}\nonumber
\\
&=& \bigl(\bigl(1-\pi_0^{*(t)}\bigr) g_1^{(t)}\bigl(\mathbf{x}_k, n_k \vert\Psi^{(t)}\bigr)
h_1^{(t)}\bigl(n_k \vert c_0^{(t)}\bigr)\bigr)
\\
&&{}\big/\bigl(\pi_0^{*(t)} g_0^{(t)}\bigl(\mathbf{x}_k, n_k
\vert\Psi^{(t)}\bigr)h_0^{(t)}\bigl(n_k \vert c_0^{(t)}\bigr)\nonumber
\\
&&{}\hspace*{7pt}+ \bigl(1-\pi_0^{*(t)}\bigr)
g_1^{(t)}\bigl(\mathbf{x}_k, n_k \vert\Psi^{(t)}\bigr)h_1^{(t)}\bigl(n_k \vert
c_1^{(t)}\bigr)\bigr),\nonumber
\\\label{3E:IP0}
\hat{I}_0^{(t)}(P_{k,i}) &\equiv& E\bigl(P_{k, i} \vert T_k=0, x_{k,i}, \Psi
^{(t)}\bigr)\nonumber
\\[-1pt]
&=& \frac{P(P_{k, i}=1, x_{k,i} \vert T_k=0, \Psi^{(t)})}{P(x_{k,i} \vert
T_k=0, \Psi^{(t)})}
\\[-1pt]
&=& \frac{(1-\pi_0^{(t)}) f_1^{(t)}(x_{k,i})}{\pi
_0^{(t)}f_0^{(t)}(x_{k,i}) + (1-\pi_0^{(t)}) f_1^{(t)}(x_{k,i})},\nonumber
\\\label{3E:IP1}
\hat{I}_1^{(t)}(P_{k,i}) &\equiv& E\bigl(P_{k, i} \vert T_k=1, x_{k,i}, \Psi^{(t)}\bigr)\nonumber
\\[-8pt]\\[-8pt]
&=& \frac{(1-\pi_1^{(t)}) f_1^{(t)}(x_{k,i})}{\pi
_1^{(t)}f_0^{(t)}(x_{k,i}) + (1-\pi_1^{(t)}) f_1^{(t)}(x_{k,i})}.\nonumber\vspace*{1pt}
\end{eqnarray}

M-step:
Now we need maximize $Q(\Psi, \Psi^{(t)})$. Since the mixing
proportions and the distribution parameters can be factorized into
independent terms, we can optimize them separately. The MLE of the
mixing proportion $\pi_0^*$ is\vspace*{1pt}
%
%e5.7 ###
%e5.6 ###
%e5.5 ###
\begin{eqnarray}
\pi_0^{*(t+1)} &=& \frac{\sum_{k=1}^N(1-\hat{T}_k^{(t)})}{N},
\\[-2pt]
\pi_0^{(t+1)}&=& \frac{\sum_{k=1}^{N}[(1-\hat{T}_k^{(t)})\sum
_{i=1}^{n_k}(1-\hat{I}_0^{(t)}(P_{k,i}))]}{\sum_{k=1}^{N}(1-\hat
{T}_k^{(t)}) n_k},
\\[-2pt]
\pi_1^{(t+1)}&=& \frac{\sum_{k=1}^{N}[\hat{T}_k^{(t)}\sum
_{i=1}^{n_k}(1-\hat{I}_1^{(t)}{(P_{k,i})})]}{\sum_{k=1}^{N}\hat
{T}_k^{(t)} n_k}.\vspace*{1pt}
\end{eqnarray}

If incorporating ancillary features of peptides, we replace
$f_j(x_{k_i})$ with $f_j(x_{k_i})\times f_j^{\mathit{nmc}}(\mathit{nmc}_{k,i})
f_j^{\mathit{ntt}}(\mathit{ntt}_{k,i})$ as in Section \ref{SS:ancillary}, where $x_{k_i}$
is the identification score, $\mathit{nmc}_{k,i}$ is the number of missed
cleavage and $\mathit{ntt}_{k,i}$ is the number of tryptic termini (with values
$s=0,1,2$). As described in Section \ref{SS:f}, $f_0 = N(\mu, \sigma
^2)$ and $f_1 = \operatorname{Gamma}(\alpha, \beta, \gamma)$. We can obtain closed
form estimators for $f_0$ as follows, and estimate $f_1$ using the
numerical optimizer $\operatorname{optimize}(\cdot$) in~R:
%
%e5.8 ###
\begin{equation}
\quad\hspace*{15pt} \mu= \frac{\sum_{k=1}^N \sum_{i=1}^{n_k} [(1-\hat{T}_k^{(t)})(1-\hat
{I}^{(t)}_0(P_{k,i})) + \hat{T}_k^{(t)}(1-\hat{I}^{(t)}_1(P_{k,i}))
]x_{k_i}}{\sum_{k=1}^N \sum_{i=1}^{n_k} [(1-\hat{T}_k^{(t)})(1-\hat
{I}^{(t)}_0(P_{k,i})) + \hat{T}_k^{(t)}(1-\hat{I}^{(t)}_1(P_{k,i}))
]},
\end{equation}
%
%e5.9 ###
\begin{eqnarray}
\quad \sigma^2 &=&\sum_{k=1}^N \sum_{i=1}^{n_k} \bigl[\bigl(1-\hat
{T}_k^{(t)}\bigr)\bigl(1-\hat{I}^{(t)}_0(P_{k,i})\bigr) + \hat{T}_k^{(t)}\bigl(1-\hat
{I}^{(t)}_1(P_{k,i})\bigr) \bigr]\nonumber
\\
&&{}\hspace*{33pt}\times {(x_{k_i}-\mu_0)^2}
\\
&&{}\bigg/ \sum_{k=1}^N \sum_{i=1}^{n_k}
\bigl[\bigl(1-\hat{T}_k^{(t)}\bigr)\bigl(1-\hat{I}^{(t)}_0(P_{k,i})\bigr)
+ \hat
{T}_k^{(t)}\bigl(1-\hat{I}^{(t)}_1(P_{k,i})\bigr) \bigr].\nonumber
\end{eqnarray}

As described in Section \ref{SS:ancillary}, we discretize NMC, which
usually ranges from 0 to 10, into states $s=0, 1, 2$, with $s=2$
representing all values $\geq2$. So the MLE of $f_0^{\mathit{nmc}}$ is
%
%e5.10 ###
\begin{equation}
f_0^{\mathit{nmc}}(\mathit{nmc}_{k,i})=\frac{w_s^{(t)}}{\sum_{s=0}^2 w_s^{(t)}},
\end{equation}
where
\begin{eqnarray}
w_s^{(t)} &=& \sum^{N}_{k=1}\sum^{n_k}_{i=1}\bigl(1-\hat{T}_k^{(t)}\bigr)\bigl(1-\hat
{I}_0^{(t)}(P_{k,i})\bigr) 1(\mathit{nmc}_{k,i}=s)\nonumber
\\[-8pt]\\[-8pt]
&&{}+ \sum^{N}_{k=1}\sum^{n_k}_{i=1}
\hat{T}_k^{(t)}\bigl(1-\hat{I}_1^{(t)}(P_{k,i})\bigr) 1(\mathit{nmc}_{k,i}=s).\nonumber
\end{eqnarray}

Similarly, the MLE of $f_1^{\mathit{nmc}}$ is
%
%e5.11 ###
\begin{equation}
f_1^{\mathit{nmc}}(\mathit{nmc}_{k,i})=\frac{v_s^{(t)}}{\sum_{s=0}^2 v_s^{(t)}},
\end{equation}
where
\begin{eqnarray}
v_s^{(t)} &=& \sum^{N}_{k=1}\sum^{n_k}_{i=1}\bigl(1-\hat{T}_k^{(t)}\bigr)\hat
{I}_0^{(t)}(P_{k,i}) 1(\mathit{nmc}_{k,i}=s)\nonumber
\\[-8pt]\\[-8pt]
&&{}+ \sum^{N}_{k=1}\sum^{n_k}_{i=1}
\hat{T}_k^{(t)} \hat{I}_1^{(t)}(P_{k,i}) 1(\mathit{nmc}_{k,i} =s).\nonumber
\end{eqnarray}

The MLE of $f_j^{\mathit{ntt}}$ takes the similar form as $f_j^{\mathit{nmc}}$, $j=0,1$,
with states $s=0,1,2$.

For $h_0$ and $h_1$, the terms related to $h_0$ and $h_1$ in $Q(\Psi,
\Psi^t)$ are
%
%e5.12 ###
\begin{equation}
\sum_{k=1}^{N} (1-\hat{T}_k)\log{h_0(n_k)}
= \sum_{k=1}^{N} (1-\hat{T}_k) \log\frac{\exp(-c_0 l_k) (c_0
l_k)^{n_k}}{n_k! (1-\exp(-c_0 l_k))},
\end{equation}
%
%e5.13 ###
\begin{equation}
\sum_{k=1}^{N} \hat{T}_k\log{h_1(n_k)}
= \sum_{k=1}^{N} \hat{T}_k \log\frac{\exp(-c_1 l_k) (c_1
l_k)^{n_k}}{n_k! (1-\exp(-c_1 l_k))}.
\end{equation}
The MLE of the above does not have a close form, so we estimate $c_0$
and $c_1$ using $\operatorname{optimize}(\cdot)$ in R.

\end{appendix}

\section*{Acknowledgments}
We thank Dr. Eugene Kolker for providing the standard mixture mass
spectrometry data, and Jimmy Eng for software support and processing
the standard mixture data.

\printaddresses

\end{document}